\documentstyle[12pt,buckow1]{article}

\begin {document}
\include{epsf}

\def\titleline{
Background Field Method and Structure of Effective Action
\newtitleline
in $N=2$ Super Yang-Mills Theories}

\def\authors{I.L.Buchbinder \1ad , B.A.Ovrut \2ad}

\def\addresses{
\1ad
Department of Theoretical Physics, Tomsk State \\
Pedagogical University, Tomsk 634041, Russia\\
\2ad
Department of Physics, University of Pennsylvania, \\
Philadelphia, PA 19104 - 6396, USA\\
and\\
Institut f\"{u}r Physik, Humboldt Universit\"{a}t zu Berlin,\\
Invalidenstra{\ss}e 110, D-10115 Berlin, Germany\\
and\\
School of Natural Sciences, Institute for Advanced Study, \\
Olden Lane, Princeton, NJ 08540, USA}

\def\abstracttext{
This paper is a brief review of background field method and some of its
applications in $N=2$ super Yang-Mills theories with a matter within
harmonic superspace approach. A general structure of effective action
is discussed, an absence of two-loop quantum corrections to first
non-leading term in effective action is proved and $N=2$
non-renormalization theorem in this approach is considered.}

~\hfill hep-th/9802156

\makefront

$N=2$ supersymmetric field theories have attracted much attention due
to significant progress in understanding their quantum aspects. Modern
interest to such theories was inspired by seminal papers by Seiberg and
Witten \cite{sw} where exact instanton contribution to low-energy
effective action has been found. This result has demonstrated once more
the wonderful features of the above theories and led to forming a
research directions associated with study a general structure of
effective action in $N=2$ super Yang-Mills theories.

An adequate description of quantum $N=2$ supersymmetric field theories
should be based on formulating these theories in terms of unconstrained
$N=2$ superfields defined on an appropriate $N=2$ superspace. Such a
description is achieved within harmonic superspace approach
\cite{gikos}.

The background field method is a powerful and highly efficient tool for
study structure of quantum gauge theories (see f.e. \cite{bos}). The
attractive features of the background field method is that it allows to
preserve the manifest classical gauge invariance in quantum theory. Due
to this circumstance the background field method is very convenient
both for investigation of general properties of effective action in
gauge theories and for carrying out the calculations in concrete field
models with gauge symmetries.

This paper is a brief review of background field method for $N=2$ super
Yang-Mills theories in harmonic superspace and some of its applications
\cite{bbko,bko}.

The harmonic superspace is defined as a supermanifold parametrized by
the coordinates
$x^m_A$, $\theta^\pm_\alpha$,
$\bar{\theta}^\pm_{\dot{\alpha}}$, $u^\pm_i$ where $x^m_A$ and $u^\pm_i$
are the bosonic coordinates and $\theta^\pm_\alpha$,
$\bar{\theta}^\pm_{\dot{\alpha}}$ are the fermionic ones. The details of
denotions are given in ref.\cite{gikos}. The remarkable property of the
harmonic superspace approach is that the set of coordinates
$x^m_A$, $\theta^+_\alpha$,
$\bar{\theta}^+_{\dot{\alpha}}$, $u^\pm_i$
transforms through each other under $N=2$ supersymmetry
transformations. It allows to treat the set of these coordinates as an
independent superspace which is called an analytic subspace
\cite{gikos}. The analytic subspace is just that appropriate manifold
for formulating the $N=2$ supersymmetric field theories.

The pure $N=2$ super Yang-Mills models are described in harmonic
superspace approach by the superfield $V^{++}=V^{++a}T^a$ where
$V^{++a}$ is analytic superfield (that is it defined on analytic
subspace), $T^a$ are the internal symmetry generators and the denotion
$++$ means that this superfield $V^{++}$ has $U(1)$-charge +2. The
action for the superfield $V^{++}$ is given as follows \cite{gikos,z}
\begin{equation}
S_{SYM}[V^{++}]= \frac{1}{g^2}\int d^{12}z\sum\limits_{n=0}^\infty
\frac{(-i)^n}{n}\int du_1\dots du_n\frac{{\rm tr}V^{++}(z,u_1)\dots
V^{++}(z,u_n)}{(u^+_1,u^+_2)\dots(u^+_n,u^+_1)}
\label{1}
\end{equation}
Here $z\equiv (x^m,\theta^i_\alpha,\bar{\theta}^i_{\dot{\alpha}})$;
$i=1,2$; $(u^+_1,u^+_2)=u^{+i}_1u^+_{2i}$ and $g$ is a coupling.
This
action is invariant under the gauge transformations \cite{gikos}
\begin{equation}
\delta V^{++}=-D^{++}\Lambda-i[V^{++},\Lambda]
\label{2}
\end{equation}
where $\Lambda$ is analytic superfield parameter and the operator
$D^{++}$ was defined in ref.\cite{gikos}.

$N=2$ matter hypermultiplets are described by the analytic
superfields\\
$q^+(x_A,\theta^+,\bar{\theta}^+,u^\pm)$ or
$\omega(x_A,\theta^+,\bar{\theta}^+,u^\pm)$. The corresponding actions
have the forms
\begin{equation}
S_q[\stackrel{\smile}{q}^+,q^+]=\int d\zeta^{(-4)}du
\stackrel{\smile}{q}^+\nabla^{++}q^+
\label{3}
\end{equation}
and
\begin{equation}
S_\omega[\omega]=\int d\zeta^{(-4)}du
(\nabla^{++}\omega)(\nabla^{++}\omega)
\label{4}
\end{equation}
with $\nabla^{++}=D^{++}+iV^{++}$ and $d\zeta^{(-4)}$ be analytic
measure \cite{gikos}. Action $S_{SYM}+S_q+S_\omega$ describes
interacting system of super Yang-Mills fields and $q^+$ and $\omega$
hypermultiplets.

To construct effective action $\Gamma[V^{++}]$ depending on $V^{++}$ we
split the superfield $V^{++}$ into background $V^{++}$ and quantum
$v^{++}$ superfields, $V^{++}\rightarrow V^{++}+gv^{++}$. The gauge
transformations (\ref{2}) can be realized as background gauge
transformations $\delta V^{++}=-\nabla^{++}\Lambda$,
$\delta v^{++}=+i[\Lambda,v^{++}]$ and as quantum gauge transformations
\begin{equation}
\begin{array}{rcl}
\delta V^{++}&=&0\\
\delta v^{++}&=&-\displaystyle\frac{1}{g}\nabla^{++}\Lambda-
i[v^{++},\Lambda]
\end{array}
\label{5}
\end{equation}
where $\nabla^{++}\Lambda=D^{++}\Lambda+i[V^{++},\Lambda]$. It is worth
to point out here that the form of background - quantum splitting and
corresponding background and quantum gauge transformations are
absolutely analogous to the conventional Yang-Mills theory but not to
$N=1$ super Yang-Mills theory (see f.e.\cite{bk})

To quantize a gauge theory within background field method one should
fix only quantum gauge transformations (\ref{5}). We introduce the
gauge fixing functions in the form
\begin{equation}
{\cal F}^{(4)}=\nabla^{++}v^{++}
\label{6}
\end{equation}
and apply Faddeev-Popov procedure. As a result we obtain effective
action $\Gamma[V^{++}]$ in the form (see the details in
ref.\cite{bbko}).
\begin{equation}
e^{i\Gamma[V^{++}]}=e^{iS_{SYM}[V^{++}]}\int{\cal D}v^{++}{\cal D}
b{\cal D}c{\cal D}\phi{\cal D}\stackrel{\smile}{q}^+{\cal D}q^+{\cal D}
\omega {\rm Det}^{1/2}(\stackrel{\frown}{\Box})e^{iS_{total}}
\label{7}
\end{equation}
where
\begin{eqnarray}
S_{total}[v^{++},b,c,\phi,\stackrel{\smile}{q}^+,q^+,\omega,V^{++}]&=&
S_2[v^{++},b,c,\phi,\stackrel{\smile}{q}^+,q^+,\omega,V^{++}]+
\nonumber\\
&+&S_{int}
[v^{++},b,c,\stackrel{\smile}{q}^+,q^+,\omega,V^{++}]
\label{8}
\end{eqnarray}
Here $S_2$ plays a role of action of free theory
\begin{eqnarray}
S_2[v^{++},b,c,\phi,\stackrel{\smile}{q}^+,q^+,\omega,V^{++}]=
-\frac{1}{2}\int d\zeta^{(-4)}du\,{\rm tr}\,v^{++}\stackrel{\frown}
{\Box}v^{++}- \nonumber\\
-\int d\zeta^{(-4)}du\,{\rm tr}\,(\nabla^{++}b)(\nabla^{++}c)-
\frac{1}{2}\int d\zeta^{(-4)}du\,{\rm tr}\,(\nabla^{++}\phi)
(\nabla^{++}\phi)+ \nonumber\\
+\int d\zeta^{(-4)}du\,\stackrel{\smile}{q}^+\nabla^{++}q^+
+\int d\zeta^{(-4)}du\,(\nabla^{++}\omega)(\nabla^{++}\omega)
\label{9}
\end{eqnarray}
The action $S_{int}$ describes the interactions
\begin{eqnarray}
&S_{int}[v^{++},b,c,\stackrel{\smile}{q}^+,q^+,\omega,V^{++}]=
-\int d^{12}z\sum\limits_{n=2}^\infty\displaystyle\frac{(-ig)^{n-2}}{n}
\int du_1 \dots du_n \times&\nonumber\\
&\times\displaystyle\frac{{\rm tr}\,v^{++}_\tau(z,u_1)\dots
v^{++}_\tau(z,u_n)}
{(u^+_1,u^+_2)\dots(u^+_nu^+_1)}+\int d\zeta^{(-4)}du
\stackrel{\smile}{q}^+V^{++}q^+ +&\nonumber\\
&+\int d\zeta^{(-4)}du(\nabla^{++}\omega v^{++}\omega+v^{++}\omega
\nabla^{++}\omega+(v^{++}\omega)(v^{++}\omega))&\\
&v^{++}_\tau=e^{-i\Omega}v^{++}e^{i\Omega}&\nonumber
\label{10}
\end{eqnarray}
Here $\Omega$ is a background bridge superfield \cite{gikos}. The
operator $\stackrel{\frown}{\Box}=\Box+$ terms depending on $V^{++}$ is
defined in ref.\cite{bbko}. The analytic superfields $b$ and $c$ are
Faddeev-Popov ghosts, the real analytic superfield $\phi$ is third (or
Nilsen-Kallosh) ghost.

The path integral (\ref{9}) for effective action $\Gamma[V^{++}]$ has
the form standard for quantum field theory. The free action $S_2$
(\ref{9}) defines the propagators of pure super Yang-Mills field,
matter fields and ghosts fields. The interaction $S_{int}$ (\ref{10})
defines the vertices. Eqs. (\ref{7}-\ref{10}) completely determine the
structure of perturbation expansion for calculating the effective
action $\Gamma[V^{++}]$ in a manifestly $N=2$ supersymmetric and gauge
invariant form.

As in conventional field theory one can suggest that the effective
action $\Gamma[V^{++}]$ is described in terms of effective Lagrangians
\begin{equation}
\Gamma[V^{++}]=\int d^4xd^4\theta d^4\bar{\theta}{\cal L}_{eff}+
(\int d^4xd^4\theta {\cal L}_{eff}^{(c)}+\,c.c.)
\label{11}
\end{equation}
where the ${\cal L}_{eff}$ and ${\cal L}_{eff}^{(c)}$ can be called a
general effective Lagrangian and chiral effective Lagrangian
respectively.

If the theory under consideration is quantized with background field
method the effective action $\Gamma[V^{++}]$ will be gauge invariant
under initial classical gauge transformations (background gauge
transformations). In this case this effective action should be
constructed only from strengths $W$ and $\bar{W}$ and their covariant
derivatives. Therefore the effective Lagrangians must have the following
general structure
\begin{eqnarray}
{\cal L}_{eff}&=&{\cal H}(W,\bar{W})+\mbox{terms depending on covariant
derivatives of}\ W\ \mbox{and}\ \bar{W} \nonumber\\
{\cal L}_{eff}^{(c)}&=&{\cal F}(W)+\mbox{terms depending on covariant
derivatives of strengths}\\
&&\mbox{and preserving chirality}\nonumber
\label{12}
\end{eqnarray}

The term ${\cal F(W)}$ in chiral effective Lagrangian depending only on
$W$ is called a holomorphic effective Lagrangian. This term is leading
in low-energy limit and describes vacuum structure theory. Namely
holomorphic effective Lagrangian was a main object of Seiberg-Witten
theory \cite{sw}. The term
\begin{equation}
\int d^4xd^4\theta d^4\bar{\theta}{\cal H}(W,\bar{W})
\label{13}
\end{equation}
defines a first non-leading correction to low-energy effective action
and describes an effective low-energy dynamics.

The structure of effective action (\ref{11}-\ref{12}) is turned out to
be analogous to structure of effective action depending on chiral and
antichiral superfields in $N=1$ case. To be more precise, the chiral
effective potential \cite{w,bko} in $N=1$ case is analogous to
holomorphic effective Lagrangian ${\cal F}(W)$. The first non-leading
correction ${\cal H}(W,\bar{W)}$ in $N=2$ case is analogous to
K\"{a}hlerian effective potential in $N=1$ case \cite{bkya,bko}.

The explicit calculations of ${\cal F}(W)$ and ${\cal H}(W,\bar{W)}$ in
one-loop approximation for hypermultiplets coupled to abelian gauge
superfield have been given within harmonic superspace formulation in
\cite{bbiko}. It has been shown that ${\cal F}(W)$ is obtained in the
form analogous to Seiberg one for pure $N=2$, $SU(2)$ super Yang-Mills
model \cite{s}. The ${\cal H}(W,\bar{W)}$ was given in a form of a
series in a power of $W\bar{W}$ where a first term proportional to
$(W\bar{W})^2$ is $N=2$ generalization of known
Heisenberg-Euler effective Lagrangian.

A simple consequence of the background field formulation is that there
are no quantum corrections to ${\cal H}(W,\bar{W)}$  at two loops in
the pure $N=2$ super Yang-Mills theory. All two-loop supergraphs
contributing to the effective action within background field method are
given in Fig.1

\begin{figure}[thbp]
        \centerline{\epsfbox{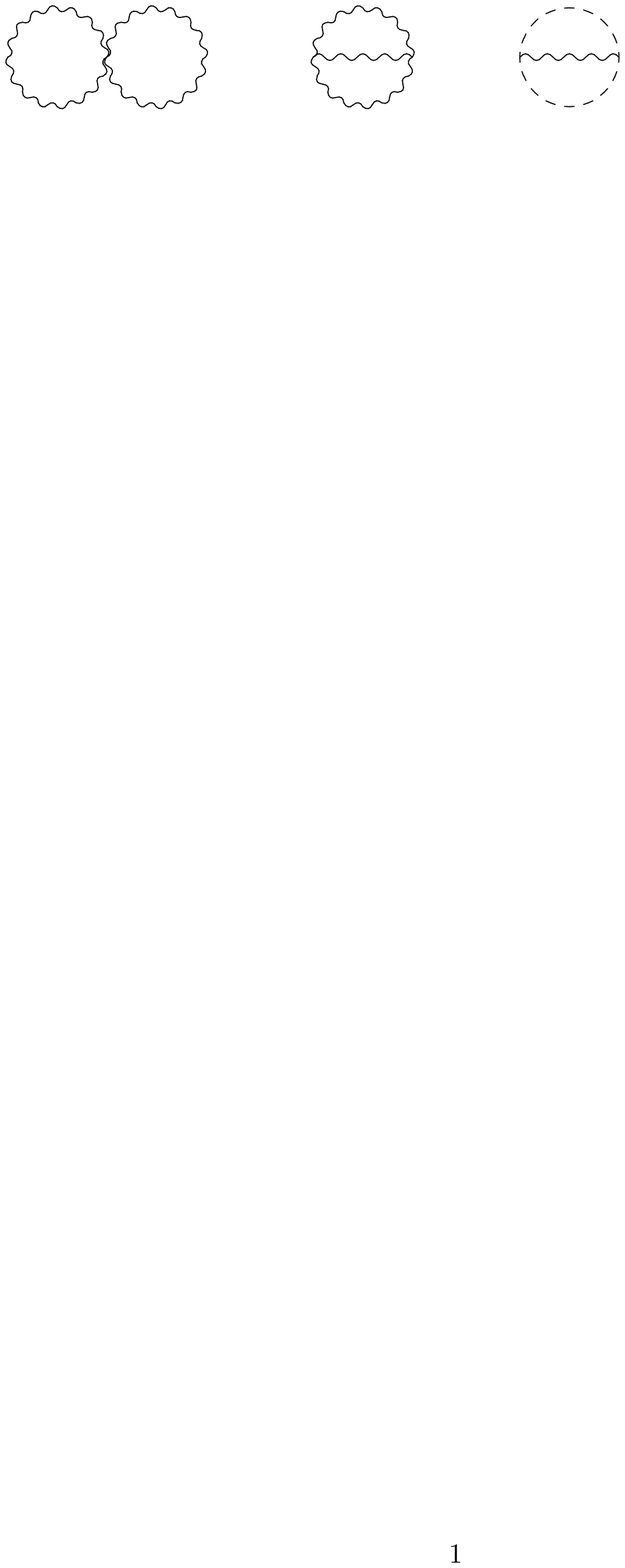}}
        \caption{Fig.1}
\end{figure}



\noindent
Here the wavy line corresponds to the super Yang-Mills propagator and
the dotted line to the ghost propagator. These propagators are defined
by the action $S_2$ (\ref{9}) and have the form
\begin{equation}
\begin{array}{rcl}
\langle v^{++}_\tau(1)v^{++}_\tau(2)\rangle&=&-\displaystyle\frac{i}
{\stackrel{\frown}{\Box}}\stackrel{\rightarrow}{({\cal D}^+_1)^4}
\{\delta^{12}(z_1-z_2)\delta^{(-2,2)}(u_1,u_2)\}\\
\langle c_\tau(1)b_\tau(2)\rangle&=&-\displaystyle\frac{i}
{\stackrel{\frown}{\Box}}\stackrel{\rightarrow}{({\cal D}^+_1)^4}
\left\{\delta^{12}(z_1-z_2)\displaystyle\frac{(u^-_1u^-_2)}
{(u^+_1u^+_2)^3}\right\}
\stackrel{\leftarrow}{({\cal D}^+_2)^4}
\end{array}
\label{14}
\end{equation}
Here $v^{++}_\tau$, $c_\tau$, $b_\tau$ and the derivatives ${\cal D}^+$
are given in so called $\tau$-frame \cite{gikos,bko} and the
distributions $\delta^{(-2,2)}(u_1,u_2)$, $(u^+_1u^+_2)^{-3}$ were
introduced in refs.\cite{gios}.

As we have noted in ref.\cite{bko}, in order to get a non-zero result
in two-loop supergraphs we should use twice the identity
$\delta^8(\theta_1-\theta_2)({\cal D}^+_1)^4({\cal D}^+_2)^4
\delta^8(\theta_1-\theta_2)=(u^+_1u^+_2)\delta^8(\theta_1-\theta_2)$
\cite{gios}. This implies that we should have 16 spinor covariant
derivatives to reduce the $\theta$-integrals over the full $N=2$
superspace to a single one. All these spinor derivatives come or from
$\stackrel{\rightarrow}{({\cal D}^+)^4}$ in the propagators (\ref{14})
or from expansion the operator $\stackrel{\frown}{\Box}^{-1}$ in a
power series of the $W$ and $\bar{W}$. After we use one
$({\cal D}^+)^4$-factor from the ghost propagator to restore the full
superspace measure, we see the propagators of both gauge and ghost
superfields have at most a single factor $({\cal D}^+)^4$. It is
evident that the number of these $({\cal D}^+)$-factors is not
sufficient to form al 16 $({\cal D}^+)$-factors we need in two-loop
supergraphs. As to a possible way to get extra $({\cal D}^+)$-factors
from $\stackrel{\frown}{\Box}^{-1}$ we observe that the spinor
covariant derivatives enter the $\stackrel{\frown}{\Box}$ always
multiplied by the derivatives of $W$ and $\bar{W}$ (see explicit form
in of $\stackrel{\frown}{\Box}$ in ref.\cite{bbko}). If we omit these
derivatives the operator $\stackrel{\frown}{\Box}$ takes the form
$\stackrel{\frown}{\Box}={\cal D}^m{\cal D}_m+\frac{1}{2}\{\bar{W},W\}$
and does not contain the spinor covariant derivatives. Therefore,
the two-loop supergraphs given in Fig.1 do not contribute to the
function ${\cal H}(W,\bar{W})$ in effective Lagrangian (\ref{12}).
It is
worth to point out that this result is a simple consequence of the
$N=2$ background field method and does not demand any direct
calculations of the supergraphs. Moreover, above result will be true
even if we take into account the two-lop matter contributions to
$\Gamma[V^{++}]$. This is almost obvious since, after restoring the
full superspace measure, the matter superfield propagator following
from action $S_2$ (\ref{9}) have effectively the same structure as the
gauge and ghost superfield propagators.

The $N=2$ background field method leads to a simple and clear proof of
the $N=2$ non-renormalization theorem. See for comparison a
consideration of problem of divergences in conventional $N=2$
superspace in ref.\cite{ww}. First of all, acting the same way as in the
case of $N=1$ non-renormalization theorem (see f.e.\cite{bk}) we can
use the $({\cal D}^+)$-factors in the propagators (\ref{14}) and in the
matter superfield propagators and restore the full superspace measure
$d^4xd^4\theta d^4\bar{\theta}$ in all vertices of all supergraphs.
Then, using the identity
$\delta^8(\theta_1-\theta_2)({\cal D}^+_1)^4({\cal D}^+_2)^4
\delta^8(\theta_1-\theta_2)=(u^+_1u^+_2)\delta^8(\theta_1-\theta_2)$,
and making integration by part we can transform any supergraph
contributing to the effective action to the form containing only a
single integral over $d^8\theta$.

Let us estimate a superficial degree of divergence for the theory under
consideration. We consider an arbitrary $L$-loop supergraph $G$ with
$P$ propagators, $N_{MAT}$ external matter legs and an any number of
gauge superfield external legs. We denote by $N_D$ the number of spinor
covariant derivatives acting on the external legs as a result of
integration by parts in the process of transformating the contributions
to a single integral over $d^8\theta$. Taking into account the
dimensions of the factors $\stackrel{\frown}{\Box}$, ${\cal D}^+$ and
the loop integrals over momenta we immediately obtain
\begin{equation}
\omega(G)=4L-2P+(2P-N_{MAT}-4L)-\frac{1}{2}N_D=-N_{MAT}-\frac{1}{2}N_D
\label{15}
\end{equation}
See the details of deriving eq.(\ref{15}) in ref.\cite{bko}. The
eq.(\ref{15}) shows that all supergraphs with external matter legs are
automatically finite. As to supergraphs with pure gauge superfield
legs, they will be finite only if some non-zero number of spinor
covariant derivatives acts on the external legs. We will show that this
is always the case beyond one loop.

Let us consider the supergraph contributions after restoring the full
superspace measure at all vertices. Then we transform these
contributions to $\tau$-frame \cite{gikos,bko}. The propagators of
gauge superfield, ghost superfields and matter superfields contain the
background field $V^{++}$ only via the $\stackrel{\frown}{\Box}$ and
${\cal D}^+$-factors, that is, only via $u$-independent connections
$A_M$ \cite{gikos}. But al connections $A_M$ contain at least one
spinor covariant derivative acting on background superfield $V^{++}$
\cite{gikos}. Therefore, each external leg must contain at least one
spinor  covariant derivative. Thus, the number $N_D$ in eq.(\ref{15})
must be greater than or equal to one. It means that $\omega(G)<0$ and,
hence, all supergraphs are ultravioletly finite beyond the one-loop
level. As to one-loop contributions to effective action they are given
in terms of functional determinants \cite{bbko,bko} and demand a
special and independent investigation.

{\bf Acknowledgements} We are grateful to our co-authors E.I.Buchbinder
and S.M. Kuzenko for collaboration and valuable discussions. We would
like to thank E.A. Ivanov for critical remarks and discussions. The work
of ILB was partially supported by the grants of RFBR, project
\symbol{242} 96-02-16017, by grant of RFBR-DFG, project \symbol{242}
96-02-00180 and by grant of INTAS, INTAS 96-0308. BAO acknowledges the
POE Contract \symbol{242} OE-AC02-76-ER-03072 and the Alexander fon
Humboldt Foundation for partial support.

\end{document}